%% file: fsi-mnc-full.tex
\documentclass[letterpaper]{mc2023}
\usepackage{tabls}
\usepackage{cites}
\usepackage{epsf}
\usepackage{appendix}
\usepackage{ragged2e}
\usepackage[top=1in, bottom=1in, left=1in, right=1in]{geometry}
\usepackage{enumitem}
\setlist[itemize]{leftmargin=*}
\usepackage{caption}
\usepackage{booktabs}
\usepackage{subcaption}
\title{A Preliminary Fluid-Structure Coupling of NekRS and MOOSE via Cardinal}
\author{%
  \textbf{A. Chaube$^{1}$\footnote{corresponding author: achaube2@illinois.edu}\ , A.J. Novak$^2$, H. Yuan$^2$, E. Merzari$^3$,}\\
  \textbf{D.R. Shaver$^2$, P.F. Fischer$^4$, C.S. Brooks$^1$}\vspace{7pt} \\
  $^1$Department of Nuclear, Plasma, and Radiological Engineering,\\
   University of Illinois Urbana-Champaign \\
  {\tt achaube2@illinois.edu}, {\tt csbrooks@illinois.edu} \vspace{6pt}\\ 
  $^2$Argonne National Laboratory, \\ 
    {\tt anovak@anl.gov}, {\tt hyuan@anl.gov}, {\tt dshaver@anl.gov} \vspace{6pt} \\ 
    $^3$Ken and Mary Alice Lindquist Department of Nuclear Engineering,\\
    Pennsylvania State University \\
    {\tt ebm5351@psu.edu} \vspace{6pt} \\
    $^4$Department of Computer Science,\\
   University of Illinois Urbana-Champaign \\
 {\tt fischerp@illinois.edu} \vspace{4pt}
}
\newcommand{\authorHead}{Chaube et al.}
\newcommand{\shortTitle}{Preliminary FSI coupling of NekRS and MOOSE in Cardinal}
\usepackage{xcolor}
\usepackage[acronym,nomain,nonumberlist,nogroupskip,nopostdot]{glossaries} 
\usepackage{siunitx}
\setacronymstyle{long-short}
\loadglsentries{acros}

\makeglossaries
\usepackage{csquotes}
\begin{document}

\maketitle
\pagestyle{fancy} \cfoot{\thepage}
\fancypagestyle{firstpage}{\fancyhead[C]{\footnotesize{\emph{
M\&C 2023 - The International Conference on Mathematics and Computational Methods Applied \\
to Nuclear Science and Engineering $\cdot$ Niagara Falls, Ontario, Canada $\cdot$ August 13 -- 17, 2023}}}
\cfoot{}}
\thispagestyle{firstpage}
\fancyhead[CE]{{\scriptsize \authorHead}}
\fancyhead[CO]{{\scriptsize \shortTitle}}
\justify
\parskip 6pt plus 1 pt minus 1 pt

\begin{abstract}
\gls{fsi} is a significant phenomenon in most nuclear reactors, causing effects such as \gls{fiv} and thermally-driven \gls{cre}. We demonstrate that Cardinal, an open-source coupling of NekRS and OpenMC to MOOSE, can be used for modelling \gls{fsi} by coupling the Tensor Mechanics Module from the \gls{moose} to NekRS's \gls{ale} solver. The solid mechanics-thermal hydraulics coupling is implemented using efficient in-memory coupling and data transfers.
 We provide a preliminary demonstration of these capabilities with a 3-D FSI benchmark for an elastic block in crossflow.
\end{abstract}
\vspace{6pt}

\glsresetall

\section{INTRODUCTION} 
\gls{fsi} has an extensive presence in disciplines such as nuclear engineering, aerodynamics, hemodynamics, and energy-systems engineering, where the interconnected dynamics of the fluid-solid interfaces significantly impact physical outcomes \cite{richter_fluid-structure_2017}. In nuclear engineering, due to the high flow-rates involved, \gls{fiv} can cause  wear in heat exchangers and reactor cores, which can have ramifications in terms of repair costs, reduced performance, and safety \cite{paidoussis_review_1983}. In \glspl{sfr}, coupling between \gls{fsi} and thermal expansion contributes to \gls{cre}, which is a complex multiphysics phenomenon that incorporates neutronics, thermal-hydraulics, and solid mechanics. Fuel deformation and the ensuing multiphysics interactions can cause fuel melting, as in the case of \gls{ebr}-I \cite{kittel_ebr-i_1958}; may induce power oscillations, one hypothesis for events observed in Ph\'enix \cite{fontaine_description_2011}; or limit burnup due to excessively warped fuel elements that cannot be removed from the core \cite{wozniak_review_2020}. For \gls{cre} in particular, Wozniak et al. \cite{wozniak_review_2020} reviewed the available literature and found that
\begin{displayquote}
{\it ``No code system currently exists which tightly and robustly couples the neutronics, thermal hydraulics, and thermal mechanical physics with enough detail to fully resolve the complex core radial expansion reactivity feedback effects.''}
\end{displayquote}
As highlighted by Merzari et al. \cite{merzari_high-fidelity_2019}, there is a need to simulate \gls{fsi}-induced phenomena in advanced reactors and their components in order to accurately assess the impact of \gls{fsi} while complementing (often costly) experiments. 



Cardinal \cite{novak_coupled_2022,merzari_cardinal_2021}, an open source\footnote{\url{https://cardinal.cels.anl.gov/}} coupling of OpenMC \cite{romano_openmc_2015} and NekRS \cite{fischer_nekrs_2021} to the \gls{moose}  \cite{permann_moose_2020}, can perform complex multiphysics simulations by leveraging existing solvers that have been verified and validated. Its modular design affords more flexibility than monolithic \gls{fsi} solvers as it can incorporate additional physics with relative ease. Cardinal also offers several computational advantages, such as in-memory coupling, distributed parallel meshes, and the construction of mesh mirrors to obviate the need for strictly conformal meshes for interfacing in code coupling. The use of NekRS allows the use of higher order elements,
and GPU-acceleration of the computationally expensive \gls{cfd} solves. Additionally, Cardinal can also leverage NekRS's \acrfull{ale} \cite{richter_fluid-structure_2017,deville_high-order_2002} mesh solver. This is in contrast to the \gls{moose} \acrfull{fsi} module, which is currently limited to applications with negligible fluid flow and small fluid domain displacements due to its reliance on an acoustic \gls{fsi} formulation \cite{dhulipala_development_2022}. 

Because Cardinal integrates NekRS and OpenMC to \gls{moose} in a general manner, Cardinal can use MOOSE’s Tensor Mechanics module to model deformation of structural materials coupled to OpenMC neutronics and NekRS thermal hydraulics. However, as Cardinal has not yet been used for \gls{fsi} applications, a coupling framework between these modules must be implemented. This work will create such a framework that can be used for high-fidelity \gls{fsi} analyses. In the near future, we intend to couple neutronics to our \gls{fsi} solves as well. 

We are demonstrating a preliminary version of our coupling using the geometry and flow conditions from a 3D-\gls{fsi} benchmark of an elastic block in crossflow \cite{richter_goal-oriented_2012}. We discuss the coupling methodology, which is followed by a detailed description of the benchmark. The results demonstrate the \gls{fsi} coupling functionality, and the fluid and solid solutions obtained. We conclude with a discussion of priorities of future work.




\section{METHODOLOGY}
\label{sec:first}

The objective of this work is to couple NekRS to MOOSE for \gls{fsi}; this coupling is achieved through Cardinal, the open-source interface between NekRS, OpenMC, and MOOSE. 
OpenMC is not directly used in this paper, but further discussions of the OpenMC--MOOSE coupling within Cardinal can be found in the literature \cite{novak_coupled_2022}. However, we note that a companion paper \cite{novak_2023} in this conference has incorporated unstructured mesh particle tracking in Cardinal via the DAGMC library, which enables moving mesh problems for Monte Carlo radiation transport. The remainder of this section introduces the basic NeKRS--MOOSE \gls{fsi} coupling, beginning with (i) how the data transfers are achieved and then followed by (ii) a discussion on the coupled solve.

\subsection{Spatial Data Transfers}

\begin{figure}[htb!]
     \centering
        \includegraphics[width=0.32\linewidth]{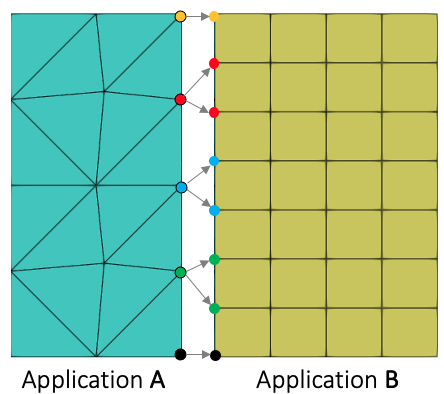}
        \caption{Example nearest-node MOOSE transfer from application $A$ to application $B$.}
        \label{fig:bdry}
\end{figure}

Data transfers between NekRS and MOOSE occur using MOOSE's {\tt Transfer} system, which provides myriad mesh-to-mesh communication options such as nearest node maps, projections, interpolations, and more. These data transfers are designed to be independent of the physical interpretation (e.g. heat flux, traction, neutron flux, ...) of the \gls{dofs} being communicated between meshes. In addition, MOOSE's {\tt Transfer} system does not mandate conformal meshes among the individual ``single physics'' codes -- each solver can use a unique mesh. An example of a boundary nearest node transfer in MOOSE between two physics applications is shown in Figure \ref{fig:bdry}.

In order for the MOOSE framework to send data in/out of NekRS, Cardinal copies NekRS's spectral element \gls{cfd} mesh into a ``mesh mirror'' of type {\tt MooseMesh} (MOOSE's unstructured mesh class). Because the mesh mirror is simply a copy of the NekRS mesh, reading/writing \gls{dofs} between the mirror and NekRS's internal representation of solution vectors is straightforward because mappings between element and node IDs are structured.

\begin{figure}[htb!]
     \centering
        \includegraphics[width=\linewidth]{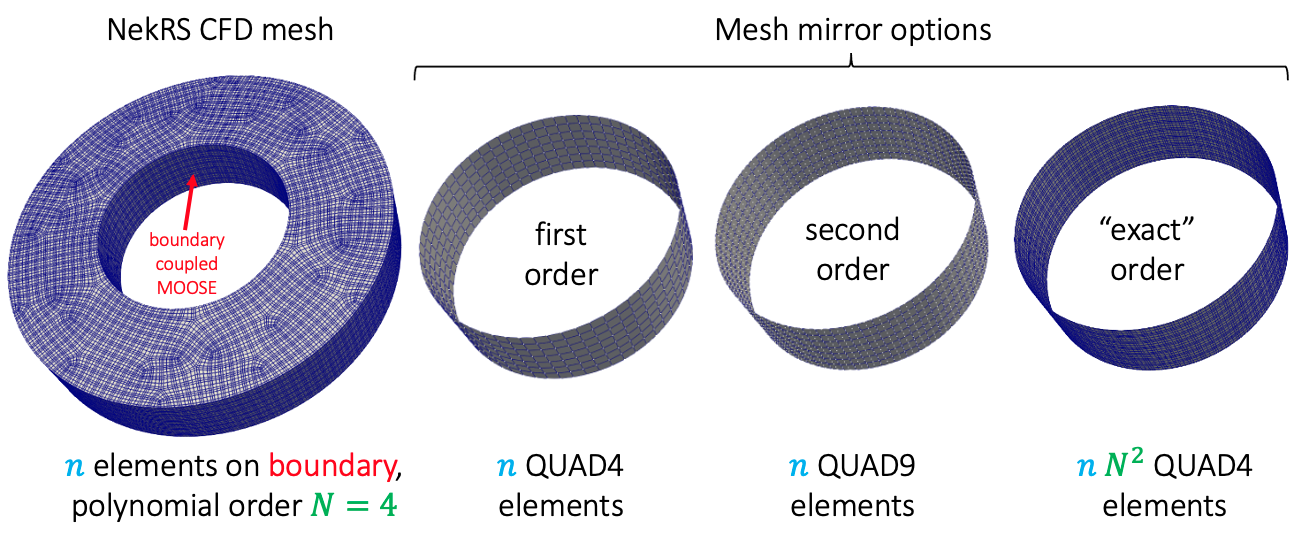}
        \caption{Example NekRS \gls{cfd} mesh (left), and different boundary mesh mirrors: (i) first-order (middle-left), (ii) second-order (middle-right), and (iii) exact (right).}
        \label{fig:bdry-mirror}
\end{figure}

For a NekRS spectral mesh of order $N$, the Cardinal mesh mirror may be either (i) a first order version of the NekRS mesh, (ii) a second order version of the NekRS mesh, or (iii) exactly identical to the NekRS mesh. Options (i) and (ii) require less memory than option (iii), at the expense of less resolution in the data transfer. Regardless of the ``order'' of the mesh mirror, 
there are two categories of mesh mirrors used. A {\it volume} mesh mirror creates a copy of the entire volumetric mesh and is useful for transferring quantities such as a volumetric heat source or fluid density, for neutronics feedback. A {\it boundary} mesh mirror is a copy of the shared boundary between applications, and is suitable for transferring heat fluxes and domain displacements.
Figure \ref{fig:bdry-mirror} shows, for illustration, a boundary ``mesh mirror'' created by Cardinal for a NekRS spectral mesh of polynomial order $N=4$.  

We note that, contrasted with previous \gls{fsi} couplings of Nek5000 to Diablo \cite{merzari_high-fidelity_2019}, that our implementation does not impose any requirements on node/element alignment between the fluid solver (NekRS) and the solid solver.
Further, our implementation may use either volume-based or boundary-based mesh mirrors. In other words, \gls{fsi} couplings of NekRS to MOOSE always communicate data through a fluid-solid interface boundary, but the \gls{fsi} physics can be optionally superimposed on top of other volume-based physics feedback (such as coupling to neutron transport).

\subsection{Coupling Methodology}


For \gls{fsi}, NekRS and MOOSE tensor mechanics are coupled via Picard iteration with a two-way data transfer to enforce continuity in boundary stress and boundary velocity. NekRS sends wall traction $t$ to MOOSE,
    \begin{equation}
        t_i=\left(-P\delta_{ij}+S_{ij}\right)n_j\ ,
    \end{equation}
where $P$ is the NekRS fluid pressure, $S_{ij}$ is the NekRS rate-of-strain tensor, and $\hat{n}$ is the unit normal on the fluid-solid interface. MOOSE sends displacement $\vec{d}$ to NekRS, which is internally converted into a boundary velocity using a first-order finite difference approximation,
\begin{equation}
\label{eq:disp}
    \vec{v}^{\ n}=\frac{\vec{d}^{\ n}-\vec{d}^{\ n-1}}{\Delta t}
\end{equation}

where $\vec{v}$ is the fluid-solid interface velocity, $n$ is the time step index, and $\Delta t$ is the time step size. This boundary velocity is then applied as a Dirichlet condition for the \gls{ale} mesh solver.

We emphasize that this \gls{fsi} implementation is preliminary, and only consists of the bare minimum requirements for the velocity and stress boundary conditions between the fluid and solid domains. Enhancements to improve stability and convergence, such as implicit partitioned predictor-corrector schemes, relaxation, and fictitious mass and damping, will be added in future work \cite{merzari_high-fidelity_2019}. 

\begin{figure}
     \centering
     \begin{subfigure}[!htb]{0.3\textheight}
         \centering
         \includegraphics[height=0.3\textheight]{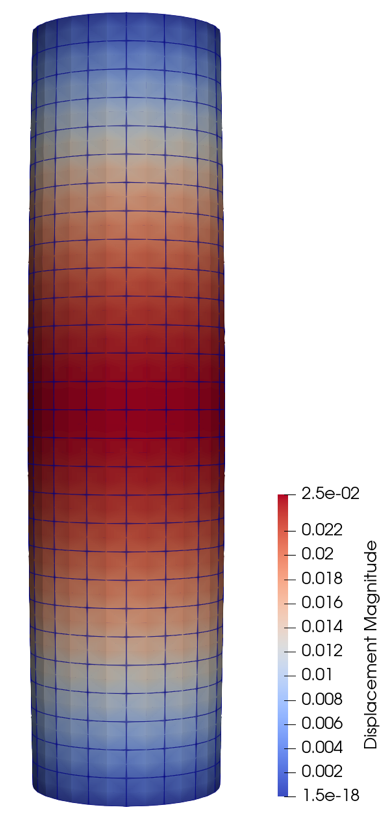}
         \caption{MOOSE mesh, with solid displacement}
         \label{fig:moose-disp}
     \end{subfigure}
     \begin{subfigure}[!htb]{0.3\textheight}
         \centering
         \includegraphics[height=0.3\textheight]{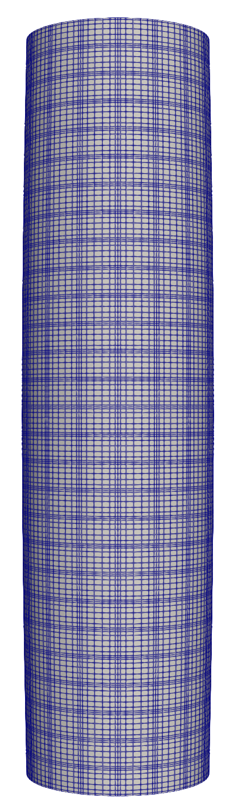}
         \caption{NekRS Mesh}
         \label{fig:nek-disp}
     \end{subfigure}
        \caption{Displacement transfer from the MOOSE mesh (left) to the spectral mesh of NekRS (right) in a test case with a swelling pipe.}
        \label{fig:swell-pipe}
\end{figure}

Initial testing of our implementation used user-prescribed function displacements within MOOSE, where a boundary displacement is provided as a generic function $f(x,y,z)$. This has allowed rapid development of \gls{fsi} data transfers without relying on a computationally intensive fully-coupled structural mechanics solve. As an example, a test case with a swelling pipe is presented in Figure \ref{fig:swell-pipe}. This case is also part of the Cardinal regression test suite.


\section{BENCHMARK}

This section presents a proof-of-concept \gls{fsi} simulation coupling NekRS \gls{cfd} with MOOSE tensor mechanics. We select a 3-D benchmark developed by Richter \cite{richter_goal-oriented_2012} due to its (i) relatively simple geometry and (ii) steady solution. Vibration-type \gls{fsi} will require additional stability enhancements in Cardinal, and are deferred to future work. 

The Richter benchmark is based on goal-oriented error estimation using the dual-weighted residual method \cite{becker_optimal_2001} for
an \gls{fsi} problem mapped entirely to the \gls{ale} domain to make the variational formulation of the \gls{fsi} problem consistent at the fluid-solid interface. 
The domain is a rectangular half-duct, with an elastic block affixed within the duct in cross-flow. The solid domain has zero-displacement
boundary conditions specified at its base, and normal displacement across the symmetric boundary is forbidden. The flow has a Reynolds number of approximately 25. Two reference values with their error are provided:
the drag exerted along the x-direction on the obstacle by the fluid and the x-displacement of the solid at a specified point.
At the interface, displacements and normal stresses are equalised.

\begin{figure}[htb!]
     \centering
     \begin{subfigure}[!htb]{\textwidth}
         \centering
         \includegraphics[width=0.65\textwidth]{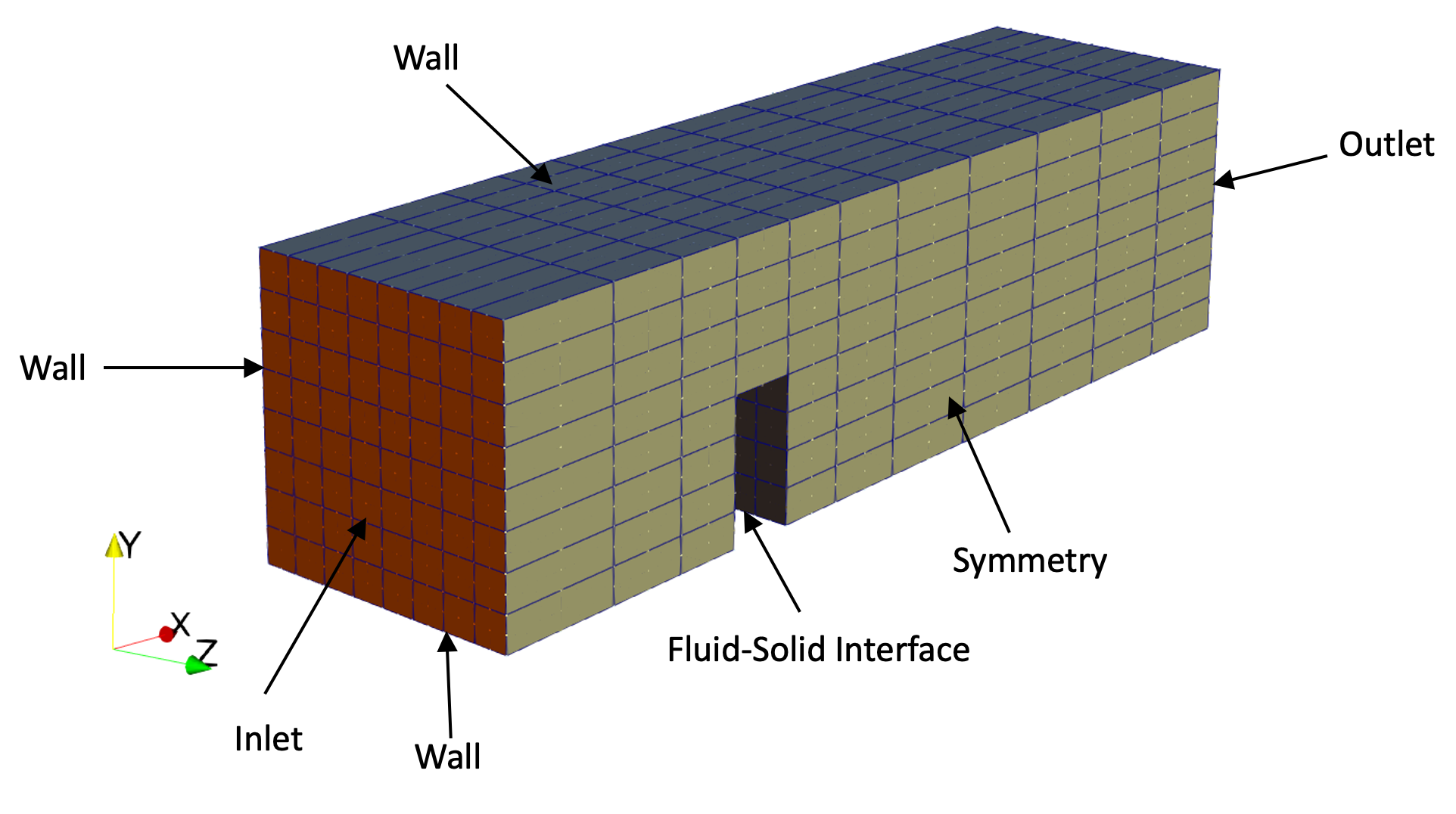}
         \caption{Fluid mesh (without Gauss-Lobatto-Legendre points) for NekRS.}
         \label{fig:nek-bc}
     \end{subfigure}

     \begin{subfigure}[!htb]{0.5\textwidth}
         \centering
         \includegraphics[width=\textwidth]{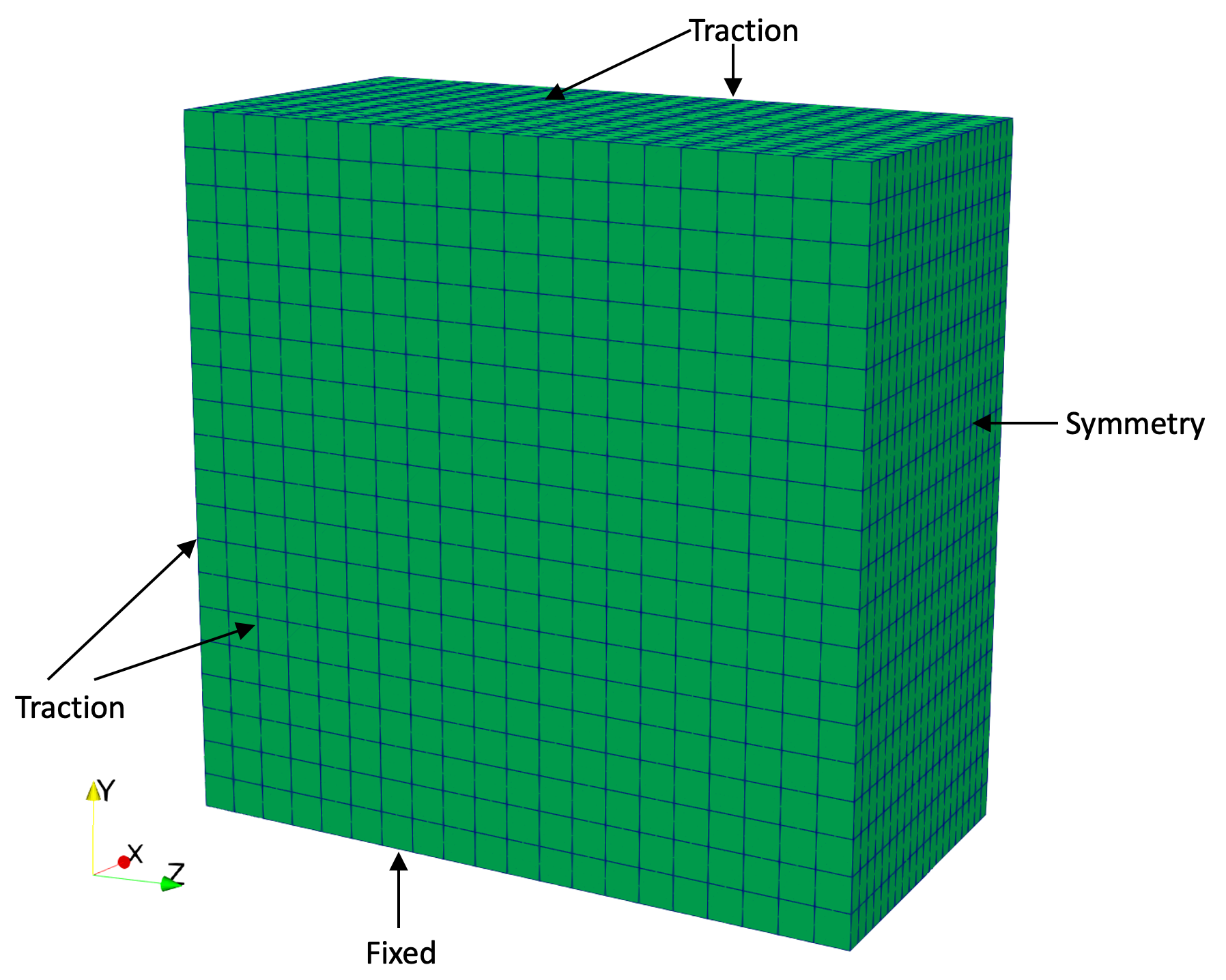}
         \caption{Second order solid mesh for tensor mechanics.}
         \label{fig:tm-bc}
     \end{subfigure}
        \caption{Meshes and boundary conditions used in the \gls{fsi} solve.}
        \label{fig:bcs}
\end{figure}

Our problem setup is the same as the benchmark; meshes and boundary conditions are shown in Fig. \ref{fig:bcs}. The fluid problem is run within NekRS using \gls{dns}, with incompressible flow and the full stress formulation enabled, on a seventh-order mesh. The \gls{moose}
Tensor Mechanics module is set up to compute finite strain for an isotropic, elastic material on a second-order mesh. The transient solves for solid mechanics are preconditioned with a high-performance hybrid multigrid preconditioner \cite{kong2020highly}. The solid block has zero displacement boundary conditions specified on the bottom, while
the stress boundary conditions are applied to the solid as traction computed within the fluid solver. The mesh displacement from the
solid is converted into mesh velocity for the \gls{ale} solver and applied appropriately. The inlet and outlet boundary conditions are from the benchmark specification. 

The initial conditions for the fluid are derived by running the standalone fluid problem without mesh movement until convergence, and applying the resulting velocity and pressure fields to the NekRS simulation as initial conditions. All data transfers occur between
the solid and the fluid problem through a second-order volume mesh mirror. While it is not strictly necessary, both solid and fluid meshes are comprised of HEX20 elements. The meshes are not conformal, as Cardinal can
interpolate data between low- and high-order meshes with sufficient accuracy. No stability enhancements such as relaxation schemes were found to be necessary, likely due to the small magnitude of displacements
involved in the benchmark problem.


\section{RESULTS}




In this section, we describe the flow and the results obtained. We note that a formal mesh refinement study has yet to be performed, and all results presented here are preliminary and should be understood as a proof-of-concept.
The simulation was run until convergence on two Intel Xeon Platinum 8260 CPUs running at 2.40GHz, which required approximately 3 hours of compute time.

As the \gls{fsi} simulation commences, the object in cross-flow experiences a sudden drag from the incompressible flow, deforming in the direction of the flow. The block oscillates as the drag fluctuates, before settling into a steady state after approximately 98 flow-throughs.

\begin{figure}[htb!]
     \centering
         \centering
         \includegraphics[height=0.3\textheight]{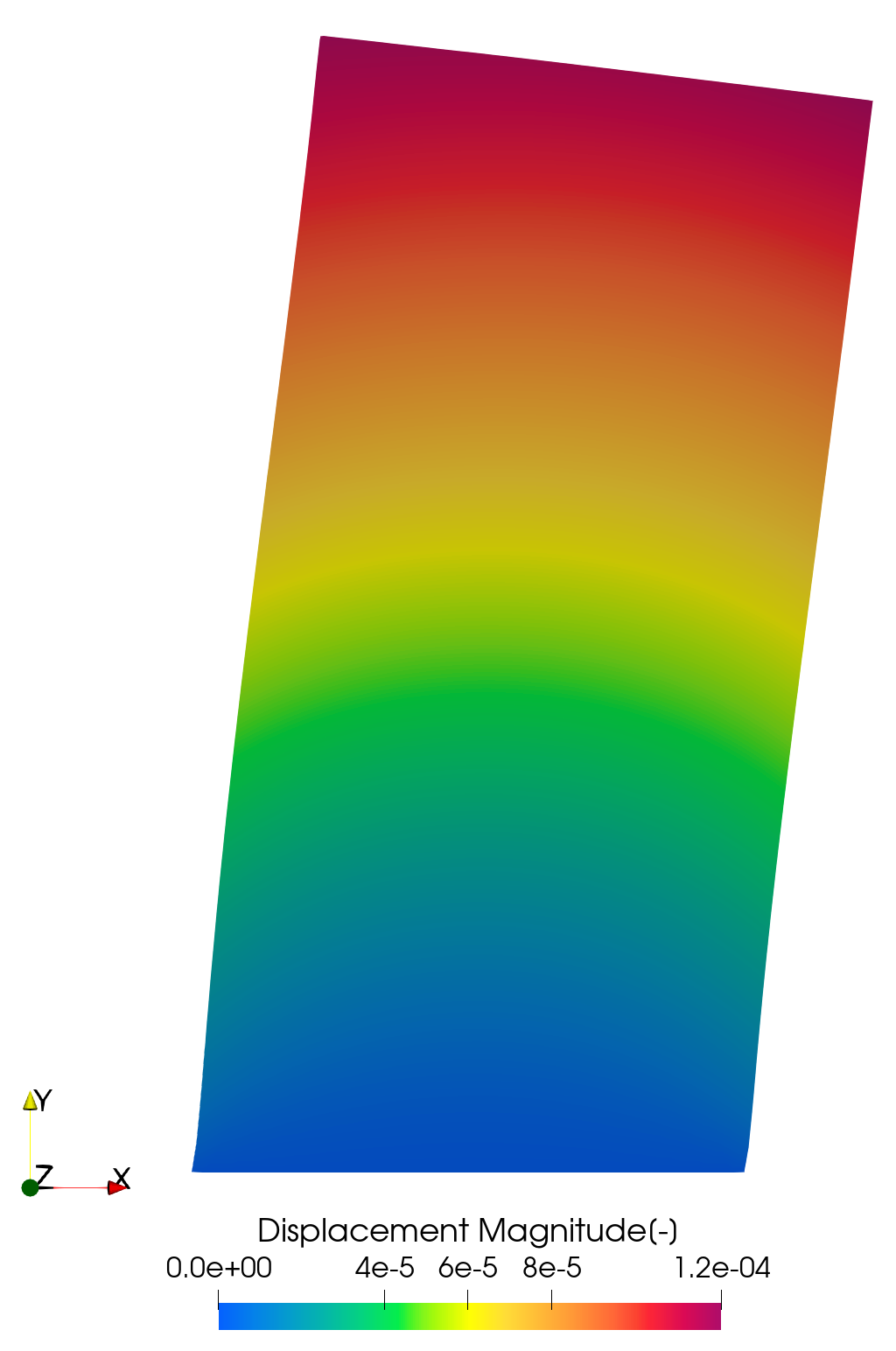}
        \caption{Displacements obtained for the \gls{fsi} problem (magnified 200x).}
                 \label{fig:displacements}
\end{figure}

\begin{figure}[htb!]
     \centering
         \centering
         \includegraphics[scale=0.5]{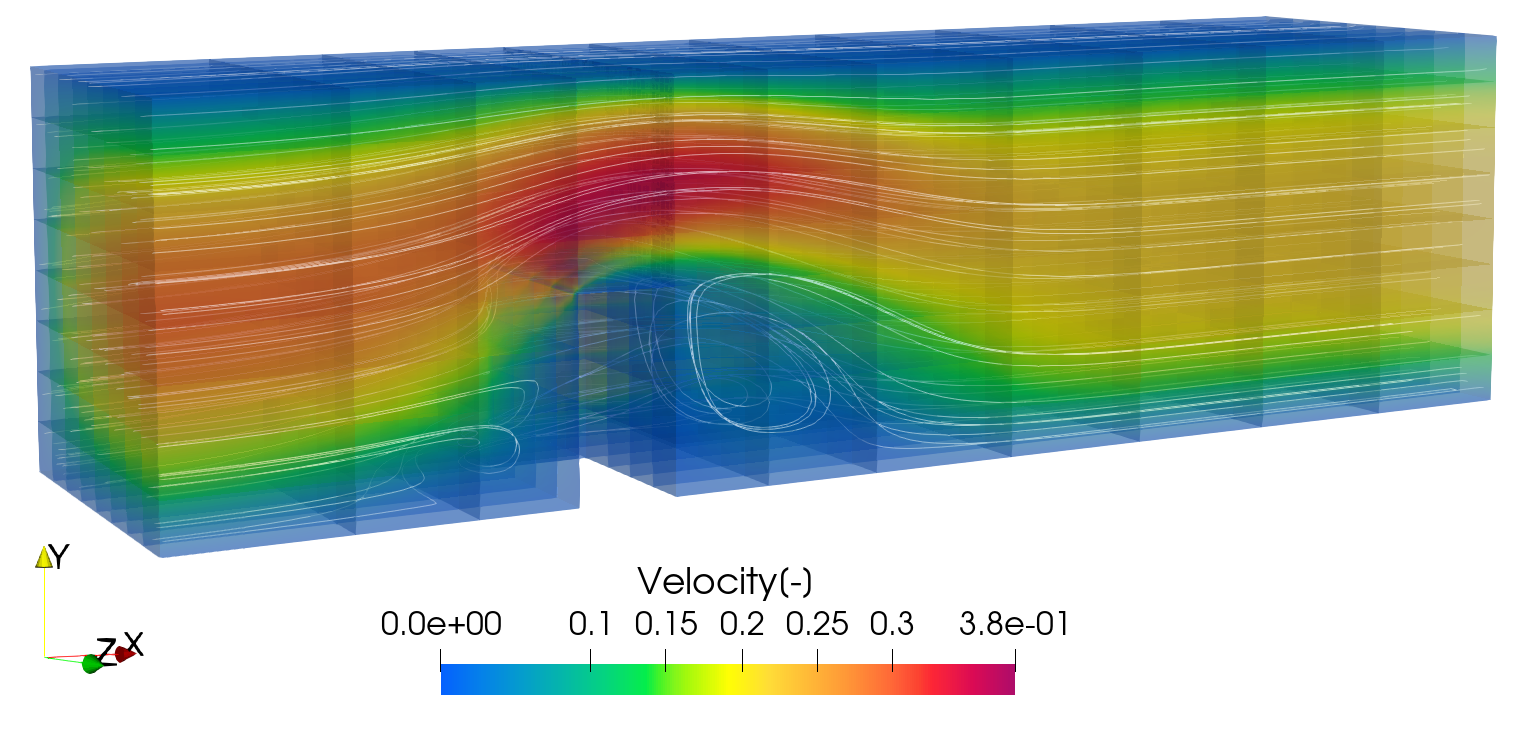}
        \caption{Velocity field from the \gls{ale}-based \gls{cfd} solve for the Richter benchmark problem.}
                 \label{fig:velocity}
\end{figure}

\begin{figure}[htb!]
     \centering
         \centering
         \includegraphics[height=0.5\textheight]{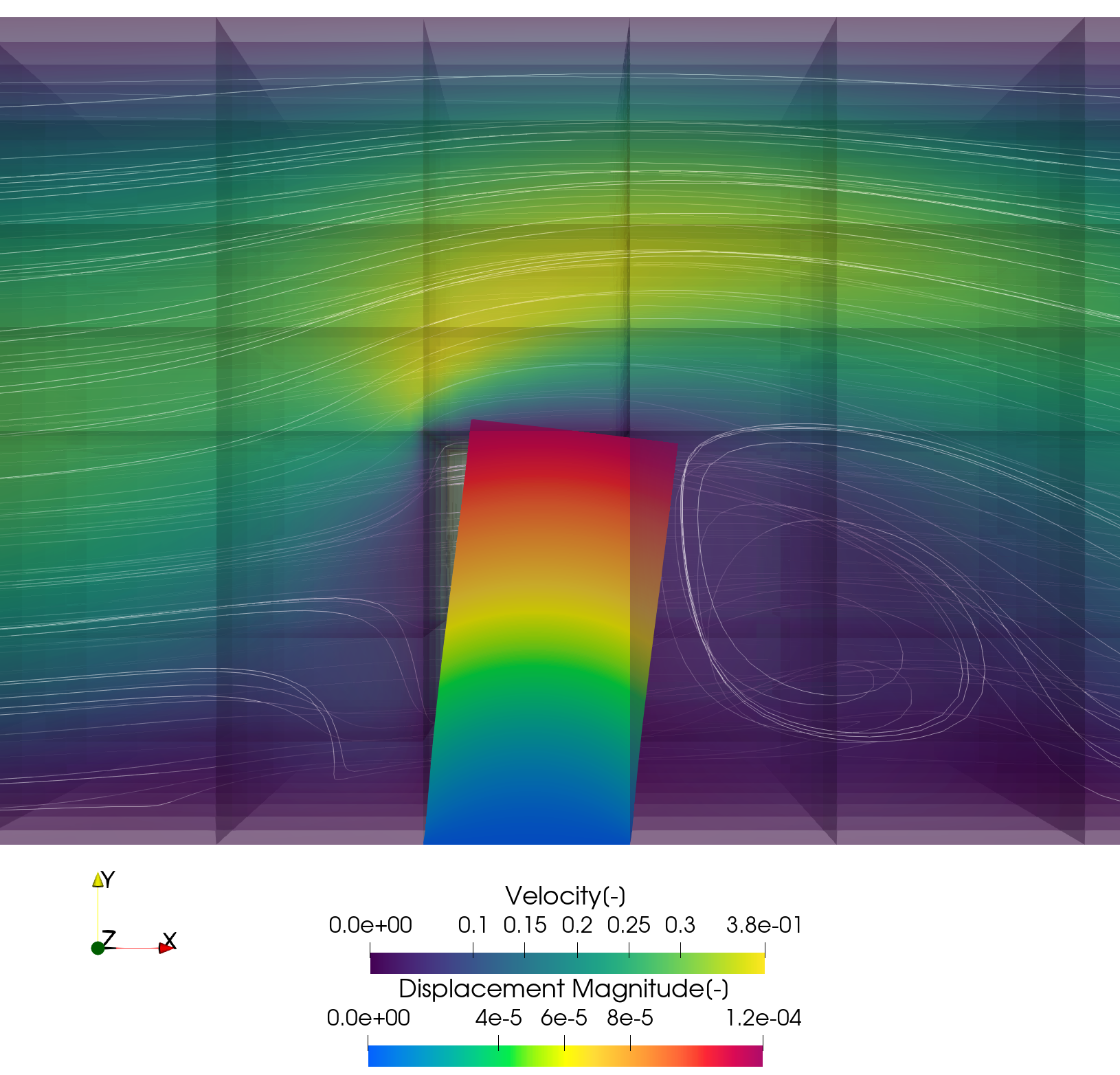}
        \caption{Solid displacements (magnified 200x) superimposed with the velocity solution on the moving fluid mesh.}
        \label{fig:velocity_with_disp}
\end{figure}

The displaced solid domain, with the displacement magnitudes is shown in Figure \ref{fig:displacements}. The velocity field with streamlines for a seventh-order spectral mesh is shown in Figure \ref{fig:velocity}. A close-up of the displaced solid and fluid solutions is shown in Figure \ref{fig:velocity_with_disp}. In order to best highlight the solid deformation, displacement of the fluid mesh is {\it not} exaggerated in either Fig. \ref{fig:velocity} or \ref{fig:velocity_with_disp}, although the mesh is indeed deforming via the \gls{ale} solver. 
While the results show comparable displacement and drag values to the benchmark, a formal mesh refinement study is needed before we can fully assess the accuracy of this particular Cardinal simulation.

\section{CONCLUSIONS}

A proof-of-concept coupling of NekRS and the \gls{moose} Tensor Mechanics module within Cardinal is demonstrated for \gls{fsi} applications. A simple block-in-crossflow problem based on a 3D \gls{fsi} benchmark was simulated, with data transfers achieved across non-conformal low-order solid mechanics and high-order \gls{cfd} domains using \gls{moose} data transfers and lower order copies of the \gls{cfd} mesh called ``mesh mirrors.'' The two-way coupling was driven by traction transfers from the fluid to the solid, and the transfer of mesh velocities from the solid to the fluid. The preliminary results show that the coupled solver is able to simulate velocity and displacement solutions for the \gls{fsi} problem. The solves are currently being performed on relatively coarse meshes for ease of software development. For our simulations, we have used a total of 921,636 degrees of freedom, which is approximately an order of magnitude less than the 7,600,775 degrees of freedom used in the benchmark itself. As next steps, we will perform a mesh refinement study and compare predictions for displacement and drag with the benchmark reference values. Large deformations, contact, and neutronics will be addressed in future work.

\section*{ACKNOWLEDGEMENTS}

The submitted manuscript has been created by UChicago Argonne, LLC, Operator of Argonne National Laboratory ("Argonne"). Argonne, a U.S. \gls{doe} Office of Science laboratory, is operated under Contract No. DE-AC02-06CH11357. The U.S. Government retains for itself, and others acting on its behalf, a paid-up nonexclusive, irrevocable worldwide license in said article to reproduce, prepare derivative works, distribute copies to the public, and perform publicly and display publicly, by or on behalf of the Government. The \gls{doe} will provide public access to these results of federally sponsored research in accordance with the DOE Public Access Plan ({\tt energy.gov/downloads/doe-public-access-plan}). This material is based upon work supported by Laboratory Directed Research and Development (LDRD) funding from \gls{anl}, provided by the Director, Office of Science, of the U.S. \gls{doe} under Contract No. DE-AC02-06CH11357.

We gratefully acknowledge use of the Bebop cluster in the Laboratory Computing Resource Center at Argonne National Laboratory.

\bibliographystyle{mc2023}
\bibliography{nourl-mnc23-fsi}

\end{document}

%% file: acros.tex

\newacronym{pwr}{PWR}{Pressurized Water Reactor}
\newacronym{lwr}{LWR}{Light Water Reactor}
\newacronym{sfr}{SFR}{Sodium-cooled Fast Reactor}
\newacronym{ale}{ALE}{Arbitrary Lagrangian-Eulerian}
\newacronym{fsi}{FSI}{Fluid-Structure Interaction}
\newacronym{fiv}{FIV}{Flow-Induced Vibration}
\newacronym{cre}{CRE}{Core-Radial Expansion}
\newacronym{cfd}{CFD}{Computational Fluid Dynamics}
\newacronym{ebr}{EBR}{Experimental Breeder Reactor}
\newacronym{rans}{RANS}{Reynold-Averaged Navier Stokes}
\newacronym{urans}{URANS}{Unsteady Reynolds-Averaged Navier-Stokes}
\newacronym{dns}{DNS}{Direct Numerical Simulation}
\newacronym{moose}{MOOSE}{Multiphysics Object Oriented Simulation Environment}
\newacronym{dofs}{DOFs}{Degrees of Freedom}
\newacronym{htgr}{HTGR}{High Temperature Gas Reactor}
\newacronym{anl}{ANL}{Argonne National Laboratory}

\newacronym{triso}{TRISO}{TRi-structural ISOtropic}
\newacronym{haleu}{HALEU}{High-Assay Low-Enriched Uranium}

\newacronym{doe}{DoE}{Department of Energy}
